\documentclass[fleqn,usenatbib]{mnras}
\usepackage{newtxtext,newtxmath}
\usepackage{amsmath}
\usepackage[T1]{fontenc}
\usepackage{subfig}
\usepackage[utf8]{inputenc} 
\usepackage{float}
\usepackage{hyperref,graphicx}
\usepackage{comment}

\title[Discovery of PSR~J1631--4722 ]{PSR~J1631--4722: The Discovery of a Young and Energetic Pulsar in the Supernova Remnant G336.7$+$0.5}
\author[Ahmad et al.]{A. Ahmad$^{1}$\thanks{e-mail: 22031320@student.westernsydney.edu.au}, S. Dai$^{2,1}$, S. Lazarevi\'{c}$^{1,2,3}$, M. D. Filipovi\'{c}$^1$, S. Johnston$^{2}$, M. Kerr$^{4}$, D. Li$^{5,6}$, 
\newauthor
C. Maitra$^{7}$, R. N. Manchester$^{2}$\\
$^1$ School of Science, Western Sydney University, Locked Bag 1797, Penrith, NSW 2751, Australia \\
$^2$ Australia Telescope National Facility, CSIRO, Space and Astronomy, P.O. Box 76, Epping, NSW 1710, Australia \\
$^3$ Astronomical Observatory, Volgina 7, 11060 Belgrade, Serbia\\
$^4$ Space Science Division, US Naval Research Laboratory, 4555 Overlook Ave SW, Washington DC 20375, USA \\
$^5$ Department of Astronomy, Tsinghua Univerisity, 30 Shuangqing Road, Beijing 100084, China \\
$^6$ National Astronomical Observatories, Chinese Academy of Sciences, Beijing 100101, China \\
$^7$ Max-Planck-Institut f\"{u}r extraterrestrische Physik, Giessenbachstra\ss e, 85748 Garching, Germany\\
}

\date{Received 2024 October 11. Revised 2024 December 6}
\pubyear{2023}
\begin{document}
\label{firstpage}
\pagerange{\pageref{firstpage}--\pageref{lastpage}}
\maketitle
\begin{abstract}
Detecting a pulsar associated with a supernova remnant (SNR) and/or pulsar wind nebula (PWN) is crucial for unraveling its formation history and pulsar wind dynamics, yet the association with a radio pulsar is observed only in a small fraction of known SNRs and PWNe. 
In this paper, we report the discovery of a young pulsar J1631--4722, associated with the Galactic SNR G336.7$+$0.5 using Murriyang, CSIRO's Parkes radio telescope. It is also potentially associated with a PWN revealed by the Rapid ASKAP (Australian Square Kilometre Array Pathfinder) Continuum Survey (RACS).
This 118\,ms pulsar has a high dispersion measure of $\sim$873\,pc\,cm$^{-3}$ and a rotation measure of --1004\,rad\,m$^{-2}$. Because of such a high DM, at frequencies below 2\,GHz, the pulse profile is significantly scattered, making it effectively undetectable in previous pulsar surveys at $\sim$1.4\,GHz. 
Follow-up observations yield a period derivative of $\Dot{P} = 3.6\times10^{-15}$, implying a characteristic age, $\tau_{c} = 33$\,kyr, and spin-down luminosity, $\Dot{E} = 1.3\times10^{36}$\,erg\,s$^{-1}$.
PSR~J1631--4722, with its high spin-down luminosity and potential link to a PWN, stands out as a promising source of the high-energy $\gamma$-ray emission observed in the region.
\end{abstract}

 
\section{Introduction}
Pulsars are highly magnetized, rapidly rotating neutron stars (NSs) that emit periodic pulses in multiple wavelengths. They are predominantly believed to be born from the core-collapse supernova explosions of massive stars, which also leave behind swept-up interstellar medium (ISM) and an expanding shell of ejected material known as supernova remnants (SNRs).
Young and energetic pulsars ($\Dot{E}\geq$10$^{36}$\,erg\,s$^{-1}$) lose a significant fraction of their rotational energy through a magnetized ultra-relativistic particle winds. These winds, when interacting with the ambient medium, generate a synchrotron-powered nebula observable from radio to beyond the X-ray bands and referred to as pulsar wind nebula \citep[PWN, for review, see][]{reynolds2017}.
A SNR enclosing a PWN is referred to as a composite-type SNR and its evolution is driven by the PWN  and occurs in several phases~\citep[e.g.,][]{2022hxga.book...77B}.
%
PWNe are believed to be a major class of Galactic sources at very high energy (VHE, E $\geq$ 0.1\,TeV)~\citep[e.g.,][]{hgps2018} and ultra-high energy (UHE, E $\geq$ 100\,TeV) $\gamma$-ray~\citep[e.g.,][]{caa+21}.
Pulsed emission from young and energetic pulsars can also be detected from GeV~\citep[e.g.,][]{sbc+19,saa+23} up to TeV~\citep[e.g.,][]{aaa+16,hess23}.
The discovery of radio pulsars associated with PWNe and SNRs holds significance in investigating the pulsar formation history, supernova explosion mechanisms, and the origin of high energy $\gamma$-ray in our Galaxy. 

Previously, many surveys have been conducted for searching young pulsars associated with SNRs and PWNe \citep[e.g.,][]{turner24,2024MNRAS.529.2443C,sett2021,straal2019,zhang2018,camilo09,lorimer1998,kaspi96,gorham96,lvm68}. To date, over 300 Galactic SNRs have been confirmed and are listed in the 2022 version of the Galactic SNR Catalogue\footnote{\url{http://www.mrao.cam.ac.uk/surveys/snrs/}} \citep{Green22}. More new Galactic SNRs are expected to be discovered, as demonstrated by the recent ASKAP SNRs detections \citep{2023MNRAS.524.1396B,2023AJ....166..149F,2024RNAAS...8..158S,2024RNAAS...8..107L}. According to the ATNF Pulsar Catalogue\footnote{\url{http://www.atnf.csiro.au/research/pulsar/psrcat/}}~\citep{manchester05}, only $\sim40$ radio pulsars have been discovered to be directly associated with known SNRs.
On the other hand, about one-third of the known PWNe (and PWN candidates) lack detection of pulsed emission in either radio or X-ray bands~\citep{ferrand12}.
\begin{figure*}
\centering
\includegraphics[width=\textwidth]{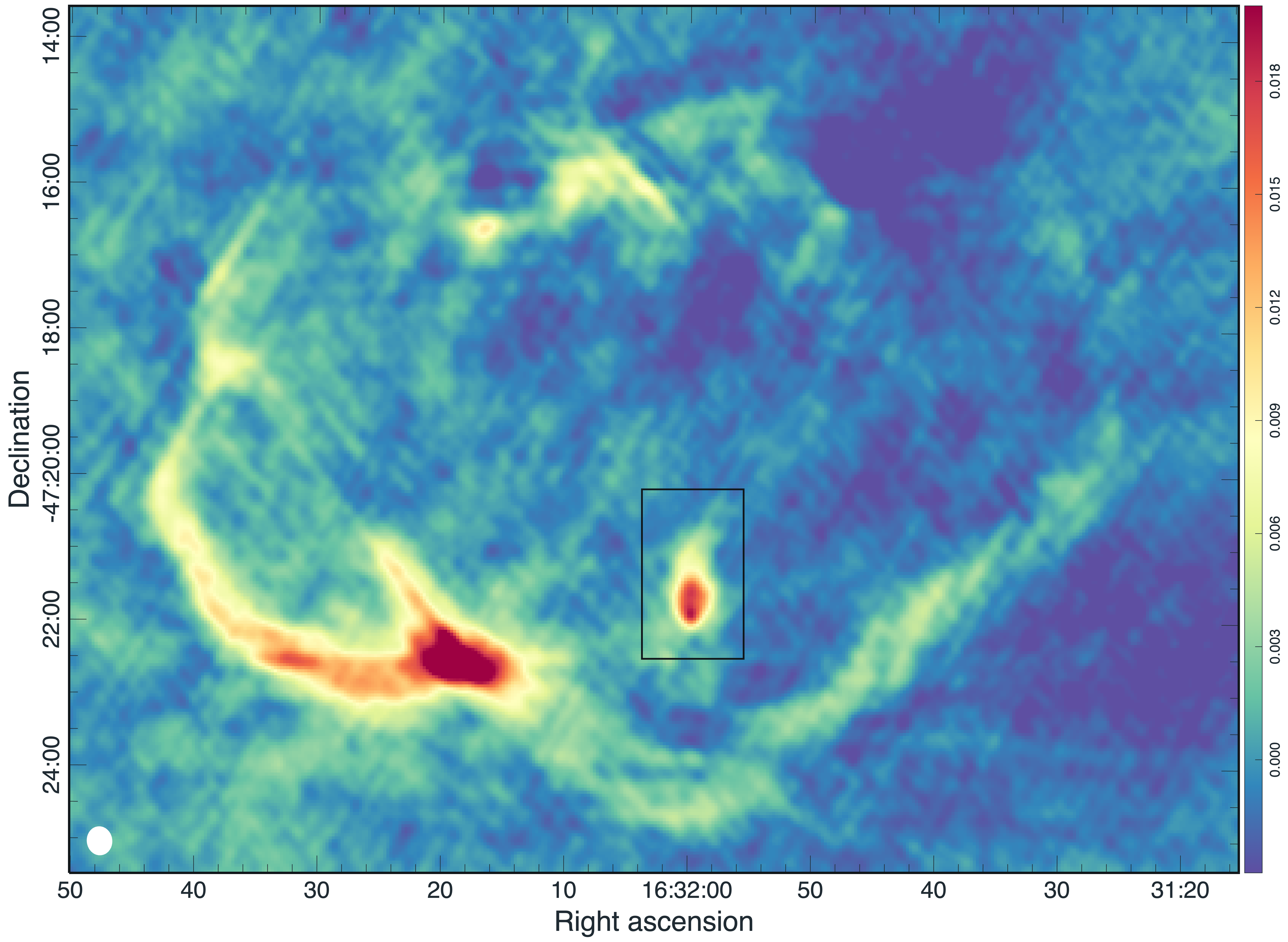}
\caption{The 888\,MHz radio continuum RACS image (SB~13601) containing the SNR~G336.7+0.5 with the putative PWN enclosed in the black rectangle. The beam size of the radio image is 14.3\,arcsec $\times$ 11.4\,arcsec and shown at the bottom left corner.}
\label{PWN}
\end{figure*}

Most SNRs and PWNe are situated in the Galactic plane, where electron density is high. Pulsar searches within these systems are susceptible to various effects, especially when the distances are large, leading to high dispersion measures (DM). Notably, radio pulses from pulsars with higher DMs undergo increased scattering due to multi-path propagation through the turbulent interstellar medium (ISM), which causes the pulse profile to become smeared.
The timescale of pulse broadening due to scattering, $\tau_{\rm s}$, scales as $\tau_{\rm s}$ $\propto$ $\nu^{\rm -4}$ \citep[e.g.,][]{krishna15}. 
Therefore, previous and ongoing pulsar surveys at mid- and low frequencies, such as the Parkes multi-beam pulsar survey~\citep{manchester2001} at 1.3\,GHz, the High Time Resolution Universe Pulsar Survey~\citep{keith10} at 1.3\,GHz, the LOFAR Tied-Array All-Sky Survey~\citep{scb+19} at 135\,MHz, the FAST Galactic Plane Pulsar Snapshot survey \citep{Su23} at 1.25\,GHz, the MPIfR-MeerKAT Galactic Plane Survey \citep[MMGPS-L;][]{padman2023} at 1.4\,GHz, and the Southern-sky MWA Rapid Two-metre pulsar survey at 150\,MHz~\citep{bsm+23} are less sensitive to highly scattered/high DM pulsars.
Targeted pulsar searches of SNRs/PWNe at higher frequencies can enhance existing efforts, as higher frequencies are less affected by scattering effects.

Thanks to the development of wide-field, deep radio continuum surveys, several highly scattered pulsars have recently been discovered at high frequencies by targeting radio continuum sources. \citet{wang23} discovered a highly scattered pulsar (PSR~J1032$‑$5804) by targeting a highly circularly polarized radio source originally identified in theASKAP Variables and Slow Transients survey~\citep{mks+21}. \citet{lfd+24} observed a bow--shock PWN (known as `Potorro') and discovered a young pulsar with the second-largest DM of known pulsars. \citet{ldj+24} discovered a highly scattered millisecond pulsar (MSP) embedded in a Galactic Centre radio filament with the largest DM and RM of any known MSP. 
With ongoing radio continuum surveys like the Evolutionary Map of the Universe (EMU) with ASKAP~\citep[EMU;][]{norris11,norris21}, more pulsar candidates are expected to be uncovered in the Galactic plane. Targeted searches at high frequencies will be the most efficient method to detect highly scattered pulsars in these regions.

In this paper, we report the discovery of a high DM, young pulsar within the SNR\,G336.7+0.5. SNR\,G336.7+0.5 is listed in the Green SNR catalogue \citep{Green22} and was initially reported in the Galactic plane survey conducted at 408\,MHz using the Molonglo Observatory Synthesis Telescope \citep[MOST;][]{shaver70}. The MOST observations at 843\,MHz with better angular resolution revealed a distinct radio shell containing an unresolved internal structure to the south of the geometric centre of SNR \citep{most96}.  
With new radio continuum images published as a part of the Rapid ASKAP Continuum Survey \citep[RACS;][]{racs20}, we identified an object with extended emission inside the SNR (see Fig.~\ref{PWN}) and then followed up with Murriyang, CSIRO's Parkes radio telescope.
%
%
In section~\ref{observations}, we summarise the discovery and follow-up observations and explain the non-detection at lower frequencies in terms of scattering. In Section~\ref{results}, we present the results including pulsar timing and polarisation properties, and pulse broadening. In Section~\ref{discuss}, we conclude our results and discuss the pulsar energetics, distance, detection of putative PWN,  as well as the potential for identifying new pulsars, particularly highly scattered ones, through the imaging domain. 

\section{Observations and Data Processing}\label{observations}
\begin{table}
\begin{center}
\caption{Measured and Derived parameters of PSR~J1631--4722 with current timing observations. }
\label{tab:psr}
\begin{tabular}{lc}
\hline
\hline
\multicolumn{2}{c}{Measured parameters}     \\
\hline
RAJ (J2000)        & 16:31:59.42(1)          \\
DECJ (J2000)       & $-$47:22:07.1(2)        \\
$\nu$ (Hz)         & 8.4232332367(8)        \\
$\dot{\nu}$ (Hz/s) & $-3.94461(6)\times10^{-12}$   \\
PEPOCH (MJD)       &  60237.254487    \\
Time span (MJD)    &  60237.26$-$60589.13   \\
DM (cm$^{-3}$\,pc) &  873.75(7)          \\
RM (rad\,m$^{-2}$) & -1004(7)  \\
Flux density at 3300\,MHz (mJy) & 0.089(2)  \\
\hline
\multicolumn{2}{c}{Derived parameters}\\
\hline
$P$ (ms)                  &  118.71925802(1) \\
$\dot{P}$ (s\,s$^{-1}$)  & $5.55963(8)\times10^{-14}$\\
$\tau_{\rm c}$ (kyr)     & 33.9 \\
$\dot{E}$ (erg/s)     &   $1.3\times10^{36}$\\
$B_{\rm s}$ (G)        & $2.6\times10^{12}$  \\
\hline
\end{tabular}
\end{center}
\end{table}
\subsection{ASKAP Observations}

The Rapid ASKAP Continuum Survey (RACS) is the first multi-frequency large sky area survey conducted using the full 36-antenna ASKAP telescope, covering the sky up to DEC $<+51^{\circ}$\citep{racs20}. The survey is being conducted at three effective central frequencies of 888, 1368, and 1655\,MHz. The integration time for each observation is $\sim$15\,minutes, reaching a typical sensitivity of $\sim$0.20--0.30\,mJy\,beam$^{-1}$ across all bands. 
The data processing was done with the ASKAPsoft package \citep{askap19}, including procedures such as calibration, flagging, and image generation for each primary beam. Subsequently, the beams were linearly mosaicked to produce a single image for each tile. PKS\,B1934--638 was used for flux density calibrations, which is the primary reference source for ASKAP \citep{hotan21}. All RACS data are publicly available through the CSIRO\,ASKAP Science Data Archive \citep[CASDA\footnote{\url{https://data.csiro.au/domain/casdaObservation}};][]{huynh20}.

In the RACS Galactic plane continuum images, we carried out a search for peculiar sources, including compact objects within known SNRs and PWNe. The field containing SNR G336.7$+$0.5 was observed within a 288\,MHz bandwidth at 1368\,MHz and 888\,MHZ frequencies, with Scheduling block (SB) identifications 22354 and 13601, respectively. The SB13601 observation used in this work was obtained on 30 April 2020. We identified a compact object with a tail-like extended emission within the SNR G336.7$+$0.5 (see Fig.~\ref{PWN}). We designated this source as RACS\,J163159.8--472157 and flagged it for further investigation.
%
\begin{figure}
    \begin{center}
    \includegraphics[width=9cm]{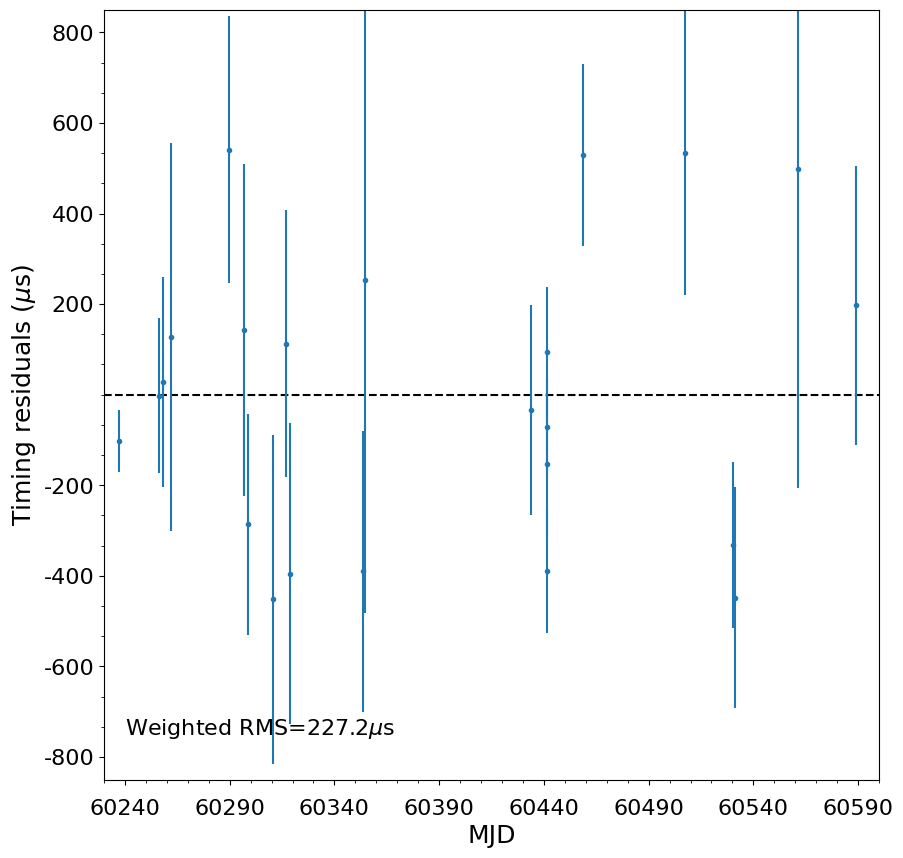}
    \caption{Timing residuals as a function of MJD for PSR J631--4722 from Murriyang obseravtions.} 
    \label{timing}
    \end{center}
\end{figure}
\subsection{Muriyang Observations}
We conducted a deep observation of RACS\,J163159.8--472157 on 20 October 2023, using Murryiyang's Ultra-Wideband Low (UWL) receiver \citep{hobs2020} in conjunction with the Medusa backend. The total integration time was 13,320\,s and only the total intensity was recorded. Data were recorded in pulsar search mode with 2-bit sampling every 64$\mu$s within 0.125\,MHz wide frequency channels, totaling 26,624 channels covering 704 to 4032\,MHz. We segmented the UWL data in two frequency subbands, with the mid-band ranging from 896\,MHz to 1920\,MHz centred at 1408\,MHz frequency, and the high-band ranging from 1984\,MHz to 3008\,MHz centred at 2496\,MHz to take into account the effects of DM smearing and scattering. The periodicity search was performed in each subband using a pulsar searching pipeline based on the \textsc{PRESTO} software package \citep{ransom01}. Periodic signals were searched within a DM range of 1--2000\,pc\,cm$^{-3}$ in the Fourier domain without accounting for any Doppler shift due to binary motion. Candidates with a signal-to-noise ratio threshold of $>8$ were folded and inspected visually. A strong pulsar candidate (hereafter designated as PSR\,J1631--4722) was identified with a period of $\sim$118\,ms and a DM of 876\,pc\,cm$^{-3}$ in the high-band data. No pulsations were detected in mid-band data below 2\,GHz.
After the initial discovery, we conducted a follow-up observation of J1631--4722 on 8 November 2023. The source was observed for $\sim$22\,minutes using UWL in the coherently de-dispersed search mode. The total intensity data were recorded with 2-bit sampling, 256$\mu$s time resolution, and 1\,MHz channel bandwidth across the entire UWL band (3328 channels). PSR\,J1631--4722 was successfully confirmed with high significance. 

A regular timing campaign of J1631--4722 was started following the confirmation as part of the Parkes young pulsar timing program~\citep{jsd+21}. The UWL system was used in the pulsar fold mode with full polarisation information and 1\,MHz frequency resolution. 
The fold mode observations were then processed and calibrated with the \textsc{PSRCHIVE} software package~\citep{hotan04} following the procedure described in \citet{sjd+21}. Each observation was visually examined using the \texttt{pazi} program to remove radio frequency interference (RFI) and then averaged in time to form an averaged pulse profile. The pulse time of arrivals (ToAs) was measured for each observation using the \textsc{pat} routine of \textsc{PSRCHIVE}. Timing analysis was carried out using the \textsc{TEMPO2} software package~\citep{hem06}.

    
\begin{figure}
    \begin{center}
    \includegraphics[width=9cm]{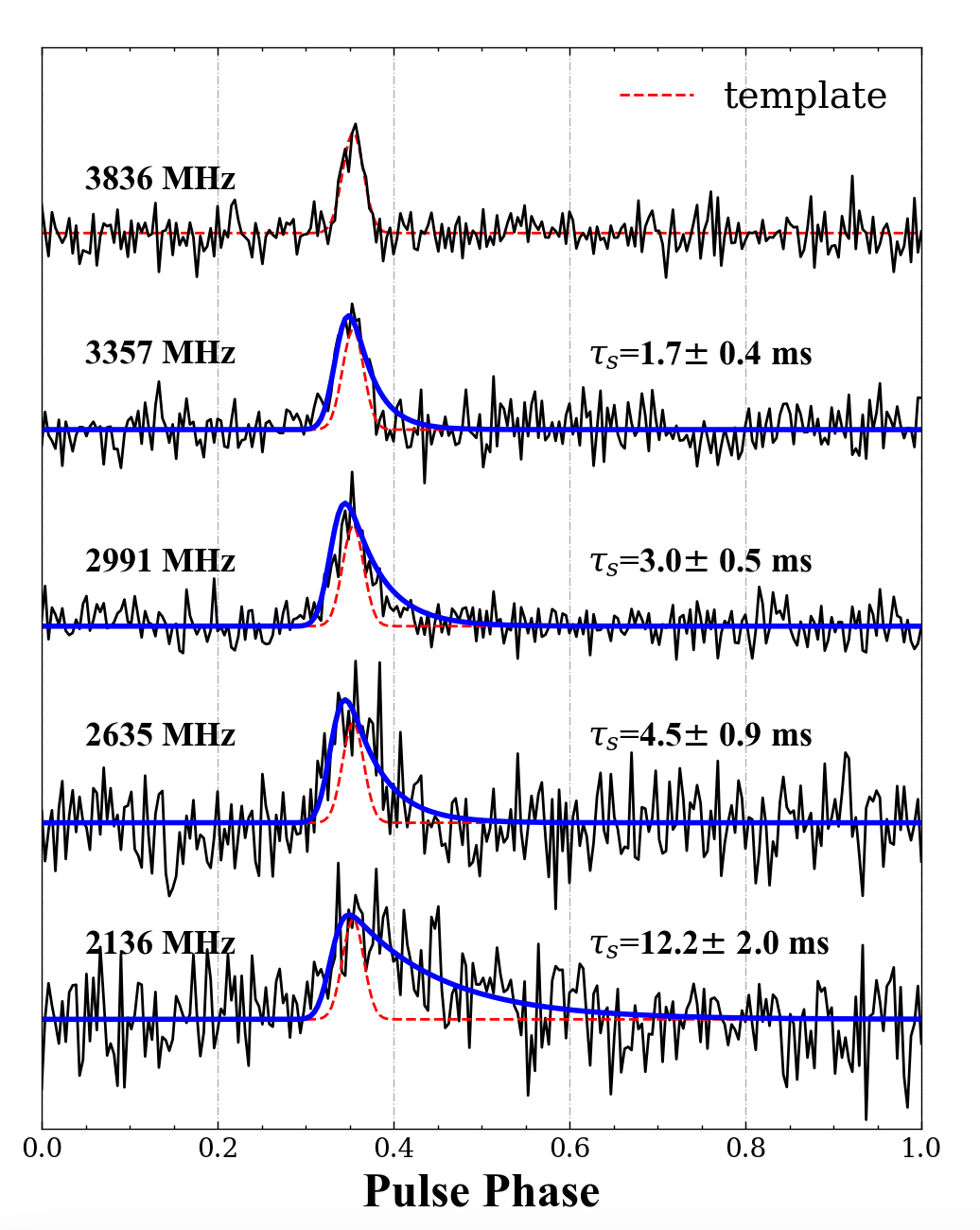}
    \caption{The multi-frequency pulse profiles of the PSR~J1631--4722 above 2\,GHz. We show the profiles at 2136, 2635, 2991, 3357, and 3836\,MHz vertically aligned from bottom to top (black), along with the assumed intrinsic pulse profile based on a fit to 3836\,MHz data (red). The blue curves represent the best-fit exponential scattering model profiles. The timescales for the scattering fits to corresponding subbands are labelled.}
    \label{broadening}
    \end{center}
\end{figure}
\section{Results}\label{results}
\subsection{Timing of PSR~J1631--4722}
A coherent timing solution of PSR~J1631--4722 has been obtained with a time span of $\sim1$\,yr. Timing residuals are presented in Fig.~\ref{timing}. In Table~\ref{tab:psr}, we listed astrometric and timing parameters of J1631--4722 together with the characteristic age $\tau_{\rm c}=P/2\Dot{P}\approx\,33.9\,\rm kyr$, spin-down power $\Dot{E}=4\pi^{2}\Dot{P}I/P^{3}\approx1.3\times10^{36}$\,erg\,s$^{-1}$, and surface magnetic field strength $B_{s}=3.2\times10^{19}\sqrt{P\Dot{P}}\approx2.6\times10^{12}$\,G. The error bars on the timing parameters were derived from \textsc{tempo2} fitting, without accounting for timing noise. The measured spin period ($P$) and spin-down rate ($\dot{P}$) place J1631--4722 within the population of young pulsars (see Fig.~\ref{ppdot}), which is consistent with its association with SNR G336.7+0.5.
%
\begin{figure}
    \begin{center}
    \includegraphics[width=9.5cm]{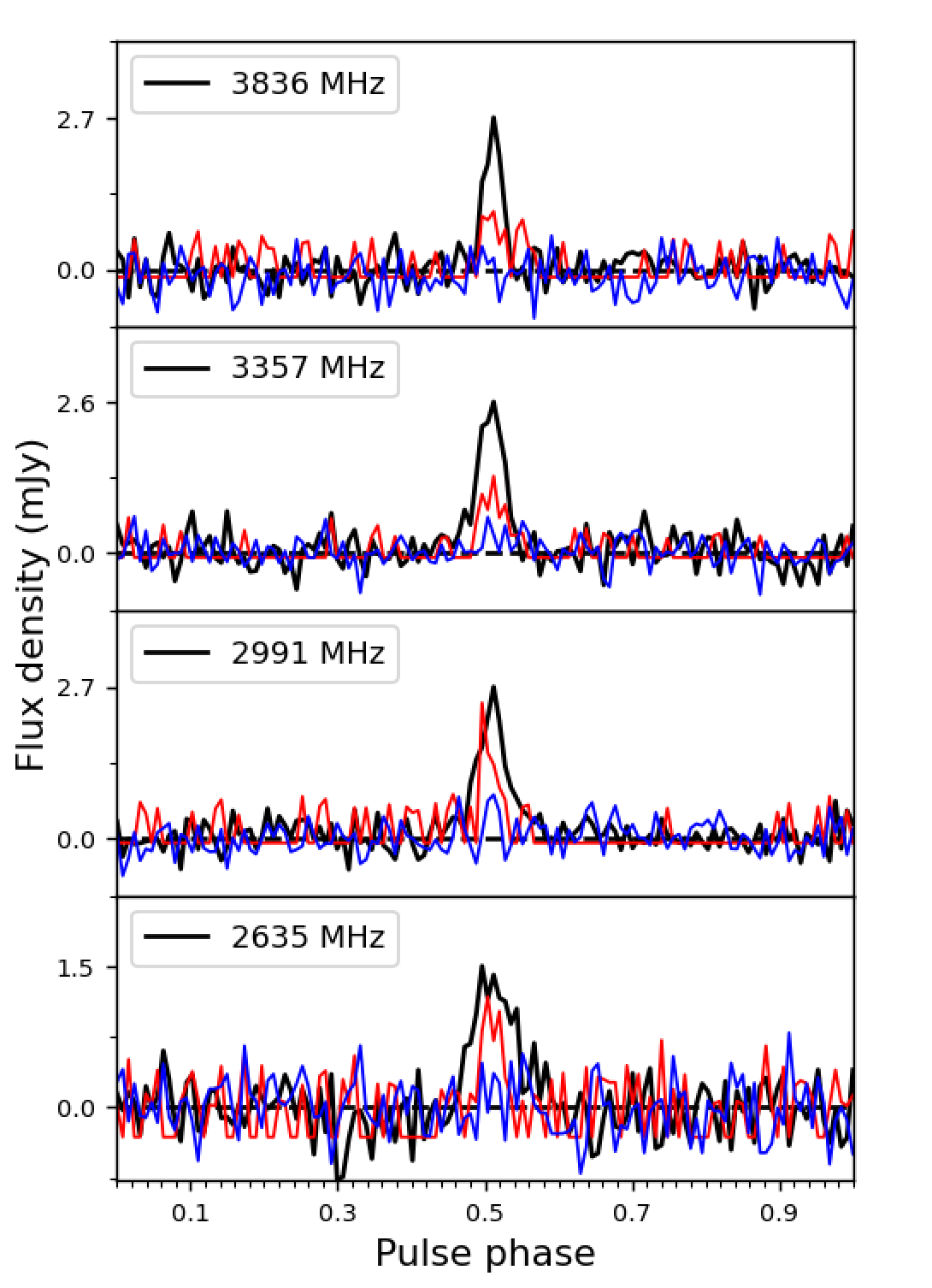}
    \caption{Averaged polarization profiles for PSR J1631--4722 in different frequency bands between 2 and 4\,GHz. The frequency-averaged profiles of the total intensity (black), linear polarisation (red), and circular
polarization (blue) are shown.}
    \label{profiles}
    \end{center}
\end{figure}
\subsection{Polarisation}
We coadded all available coherently dedispersed UWL fold mode observations into a single high-S/N pulse profile. To measure the pulsar rotation measure (RM), we extract the higher frequency subband above 2\,GHz (2368--4032\,MHz) with 1664 frequency channels. 
We determine the RM using \texttt{rmfit}\footnote{\url{http://psrchive.sourceforge.net/manuals/rmfit/}} tool of \texttt{PSRCHIVE} \citep{hotan04}. 
The RM is determined by the brute force algorithm of \texttt{rmfit} examining an RM range of $\pm$\,3000\,rad\,m$^{-2}$. At each trial RM, \texttt{rmfit} adjusts for the corresponding Faraday rotation and calculates the linear polarization $L = \sqrt{Q^{2}+U^{2}}$ across the on-pulse phase bins, where Q and U are linear Stokes parameters. Subsequently, the algorithm fits a Gaussian profile to the resulting RM spectrum to determine the peak RM. 
We found the best--fit detection with the RM of $-1004\pm7$\,rad\,m$^{-2}$. 

To examine the polarization evolution of PSR J1631--4722 with frequency, we generate multi-frequency polarization pulse profiles after averaging in time and correcting for Faraday rotation using the best-fit RM of --1004. In Fig.~\ref{profiles}, we present averaged polarization pulse profiles for four frequency sub-bands (centred at 2.6\,GHz, 2.9\,GHz, 3.3\,GHz, and 3.8\,GHz). Black, red, and blue lines represent the total intensity (I), linear polarization (L), and circular polarization (V), respectively. The baselines for the Stokes I, Q, U and V profiles have been set to zero mean. The linear polarisation L is calculated as $L = \sqrt{Q^{2}+U^{2}}$, and the noise bias in L is corrected according to equation 11 in \citet{everett2001}. The similar bias in $|\rm V|$ was corrected as described in \citet{Dai15}.

We found that PSR~J1631--4722 is linearly polarised with weak circular polarization, in particular, the highest fractional linear polarisation is observed at 3.8\,GHz frequency. The high degree of linear polarisation is a characteristic of highly energetic and young pulsars~\citep[e.g.,][]{walt+08,john+06}. At the highest frequency subband (top panel in Fig.~\ref{profiles}), the pulse profile appears somewhat narrower, with a progressively broader tail observed towards lower frequencies. An exponentially decaying tail at 2.6\,GHz (bottom panel of Fig.~\ref{profiles}) is consistent with the scattering timescale dependence on frequency as~$\sim$\,$\tau_{\rm s}$ $\propto$ $\nu^{\rm -4}$ \citep[e.g.,][]{krishna15}, indicating the evident scatter-broadening of the pulse profile below 2\,GHz.
\subsection{Pulse Broadening of PSR J1631--4722}
\begin{figure}
    \begin{center}
    \includegraphics[width=\linewidth]{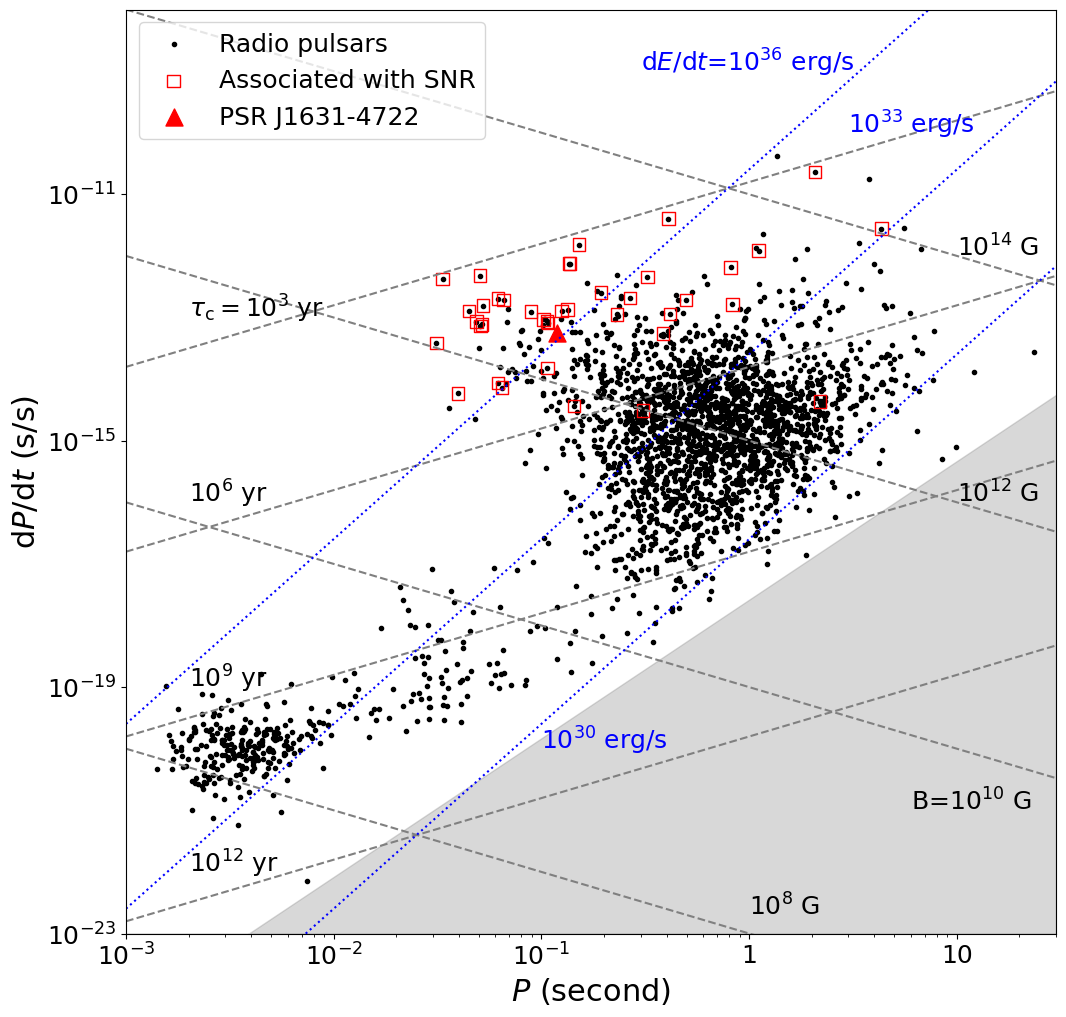}
    \caption{The spin period (P) versus the time derivative of the spin period ($\Dot{\rm P}$) for known radio pulsars. Pulsars with SNR (or potential) association according to the ATNF pulsar catalogue are marked with red squares. PSR~J1631--4722 is shown as a red triangle. We plot contours of characteristic ages, magnetic fields and spin-down luminosity based on the canonical pulsar spin-down model \citep{lorimer04}.}
    \label{ppdot}
    \end{center}
\end{figure}
\begin{figure}
    \begin{center}
    \includegraphics[width=9cm]{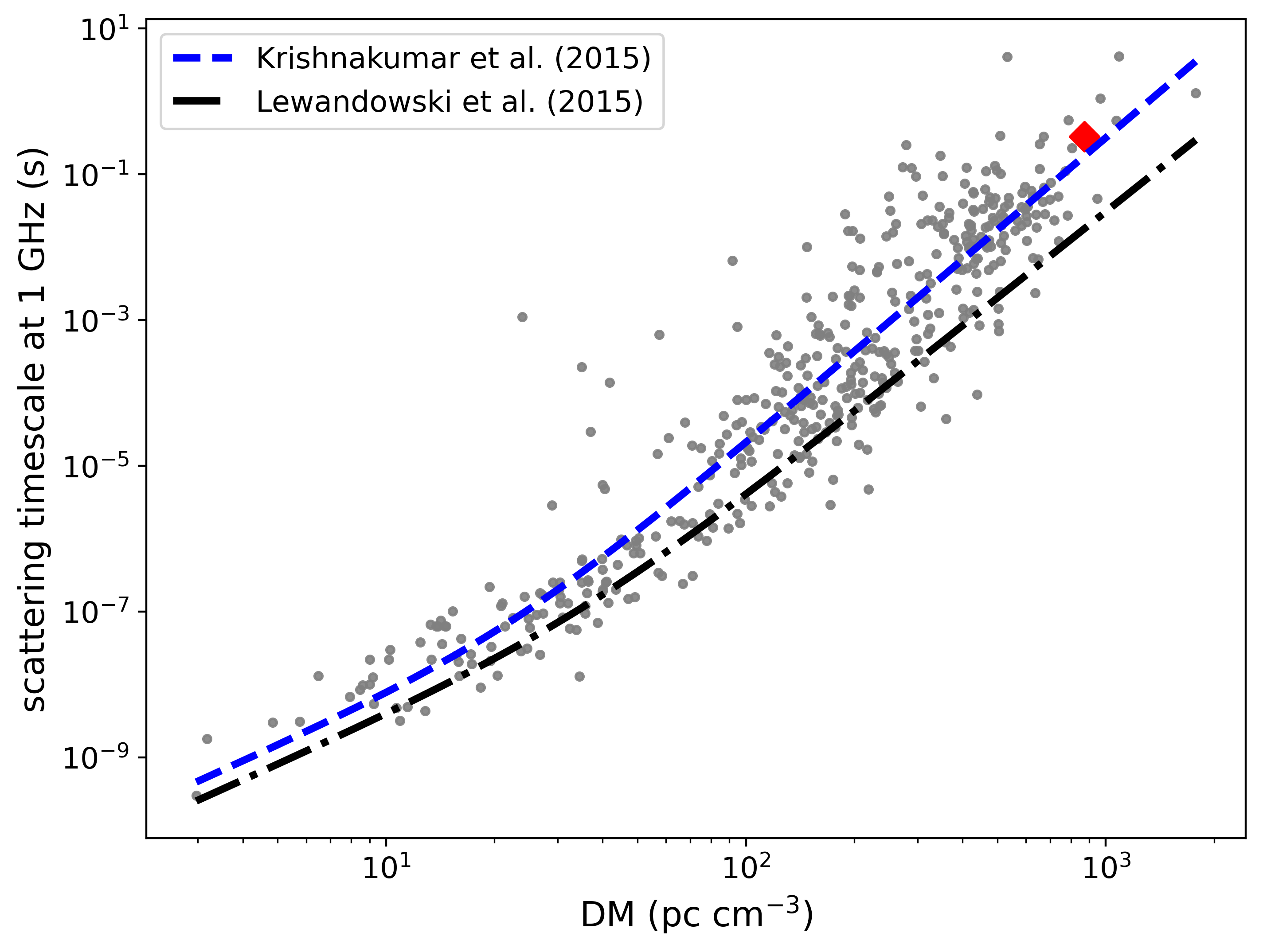}
\caption{Scattering timescale at 1\,GHz vs DM for pulsars (grey dots) in ATNF pulsar catalogue \citep{manchester05} and for PSR J1631--4722 (red diamond). The blue and black dashed lines correspond to the relation between DM and scattering timescale fitted by \citet{krishna15} and \citet{lewand15}, respectively.}  
    \label{dmscat}
    \end{center}
\end{figure}

In order to estimate the actual amount of pulse broadening, we split the data above 2\,GHz into five subbands and assume that the scattering timescales $\tau_{\rm s}$ vary with frequency as $\nu^{\rm -\alpha}$ \citep[see, e.g.,][]{oswald21}. 
We fit the pulse profile in the highest frequency subband with a 1-D Gaussian model (see Fig.~\ref{broadening}) and assume that this is the intrinsic pulse shape (top panel of Fig.~\ref{broadening}).  
We measured the scattering timescales for individual subbands independently by fitting the pulse profile convolved with an exponential tail~\citep[e.g.,][]{bdz+22}. When we fit with frequency-dependent power law, we obtain a spectral index $\alpha = \rm 4.4 \pm \rm 0.1$, consistent with the Kolmogorov turbulence in ISM \citep[e.g.,][]{bhat2004} and scattering timescale at 3\,GHz of $\rm 3.0 \pm 0.5$\,ms. This implied that a scattering timescale of 76\,ms (64$\%$ of the spin period) scaled to 1.4\,GHz frequency would smear out most of the pulsed flux for PSR J1631--4722. In Fig.~\ref{dmscat}, we plot the scattering timescales at 1\,GHz against the dispersion measure for pulsars in ATNF pulsar catalogue \citep{manchester05}. Despite the large uncertainties in the models \citep[e.g.,][]{bhat2004,lewand15,krishna15} at higher DMs (see Fig.~\ref{dmscat}), the scattering timescale of PSR~J1631--4722 is consistent with the spread of the DM--$\tau_{\rm s}$ correlation and is in agreement with the whole pulsar population.
\section{Discussion and conclusion}\label{discuss}
We report the discovery of a highly scattered, young pulsar PSR~J1631--4722 at above 2\,GHz frequency in a targeted observation with the UWL receiver of Murriyang. The pulsar is associated with the SNR G336.7+0.5 located in the Galactic plane.
Our current timing parameters give us a characteristic age of $\tau_{\rm c}\approx33.8$\,kyr; the surface magnetic field of $B_{\rm s}\approx2.6\times10^{12}$\,G; and the spin-down luminosity of $\dot{E}\approx1.3\times10^{36}$\,erg\,s$^{-1}$.
In Fig.~\ref{ppdot}, we plotted the spin period ($P$) and the spin-down rate ($\dot{P}$) for all known pulsars listed in the ATNF Pulsar Catalogue~\citep{manchester05}, and PSR J1631--4722 is represented by a red triangle. These spin parameters place PSR\,J1631--4722 in the population of highly energetic, rotation-powered pulsars.

In the vicinity of PSR~J1631--4722, there are two HGPS (H.E.S.S Galactic Plane Survey) TeV $\gamma$-ray sources HESS\,J1634--472 and J1632--478~\citep{hess} and three GeV $\gamma$-ray sources 4FGL\,J1633.0--4746, 4FGL\,J1631.6--4756, and 4FGL\,J1636.3--4731~\citep{fermi23}. In Fig.~\ref{hess}, we show the position of PSR~J1631--4722 and SNR G336.7+0.5 on the HESS $\gamma$-ray significance map together with those of Fermi sources and known pulsars in the region. While only diffused TeV $\gamma$-ray emission can be observed at the position of PSR~J1631--4722, it is within the error circle of bright and extended Fermi source 4FGL\,J1633.0--4746. Assuming that 3\% of its spin-down energy contributes to GeV $\gamma$-ray emission, we estimate a $\gamma$-ray flux of $F=0.03\times\Dot{E}/4 \pi d^{2}\approx6.68\times10^{-12}$\,erg\,s$^{-1}$\,cm$^{-2}$ at a distance of $d=7$\,kpc, which is $\sim3\%$ of the flux of 4FGL\,J1633.0--4746. 
%
\begin{figure}
    \begin{center}
    \includegraphics[width=9cm]{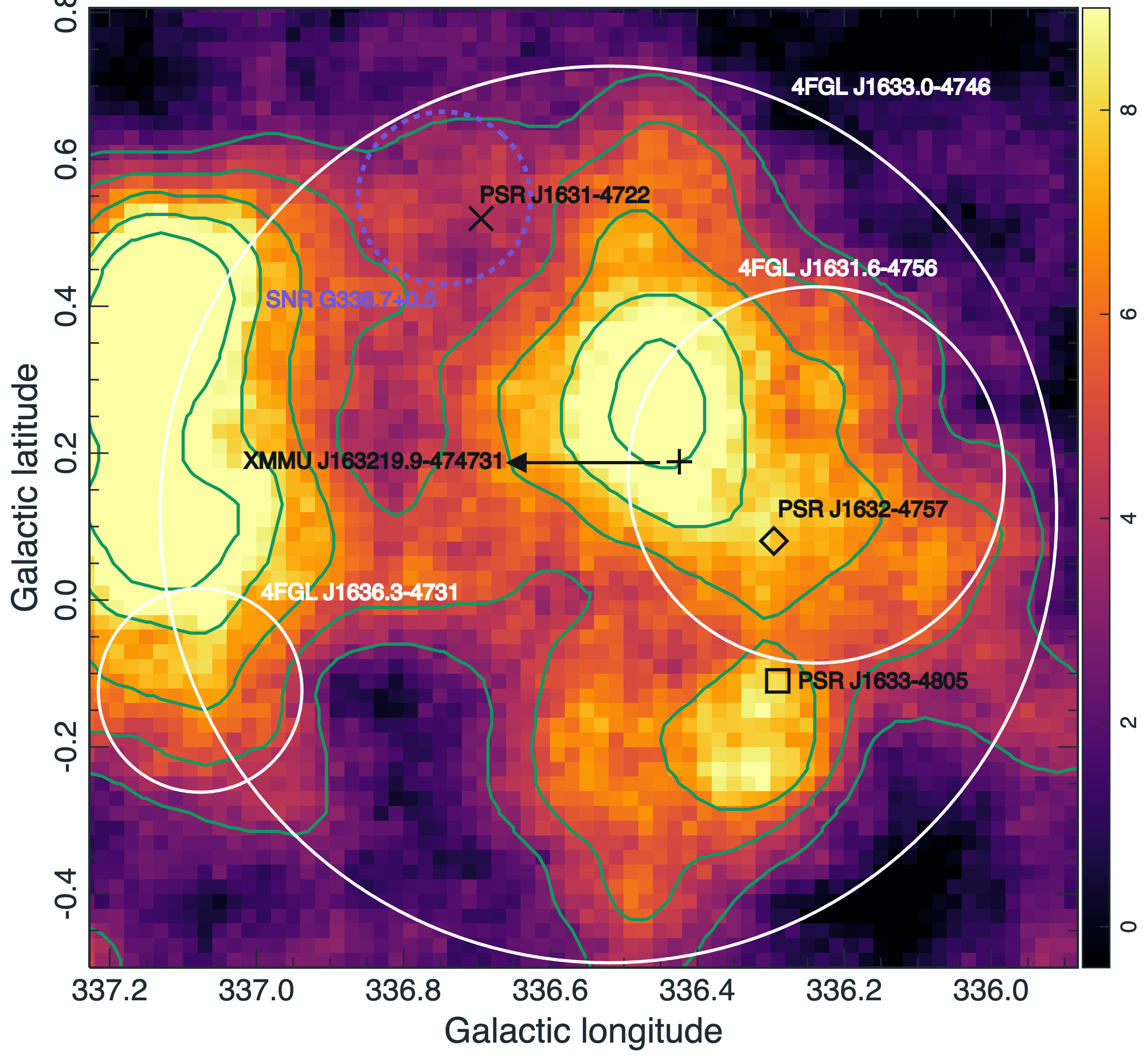}
    \caption{TeV $\gamma$-ray significance map of HESS\,J1632--478 and HESS\,J1634-472 (towards left), along with plausible counterparts, which are found to be spatially coincident on HGPS sources. SNR\,G336.7+0.5 is indicated by the purple dashed circle. PSRs\,J1631--4722, J1632--4757, and J1633--4805 are indicated by the black cross, diamond, and square, respectively. XMM\,J163219.9--474731 is an extended X-ray source close to the centroid of HESS\,J1632-4757 \citep{balbo2010} and marked by plus sign. Fermi-LAT DR4 sources are represented by white circle and accordingly labelled. TeV $\gamma$-ray emission for 3--10$\sigma$ is shown by the solid green contours. The FoV is $\sim$1.3$^{\rm o}$$\times$1.3$^{\rm o}$.}
    \label{hess}
    \label{hess}
    \end{center}
\end{figure}
It remains uncertain whether PSR~J1631--4722 contributes significantly to the observed $\gamma$-ray emissions in the region. More accurate distance measurement to the pulsar is crucial to addressing this question. Our current one-year timing baseline is insufficient to accurately predict the pulsar's spin parameters for folding high-energy data previously collected by \textit{HESS} and \textit{Fermi}. Like other young pulsars, J1631$-$4722 may also experience glitches~\citep[e.g.,][]{els+11,ymh+13,ljd+21,bsa+22}, making it very difficult to fold past datasets. Therefore, continued timing observations of J1631$-$4722 are crucial for tracking its long-term spin parameter evolution and precisely determining its astrometric properties, enabling future searches for high-energy pulsations.
Despite these uncertainties, PSR~J1631--4722 stands out as the most energetic pulsar in the region, with a spin-down luminosity far exceeding those of PSRs~J1632--4757 and J1633--4805.

\begin{figure}
    \begin{center}
    \includegraphics[width=9cm]{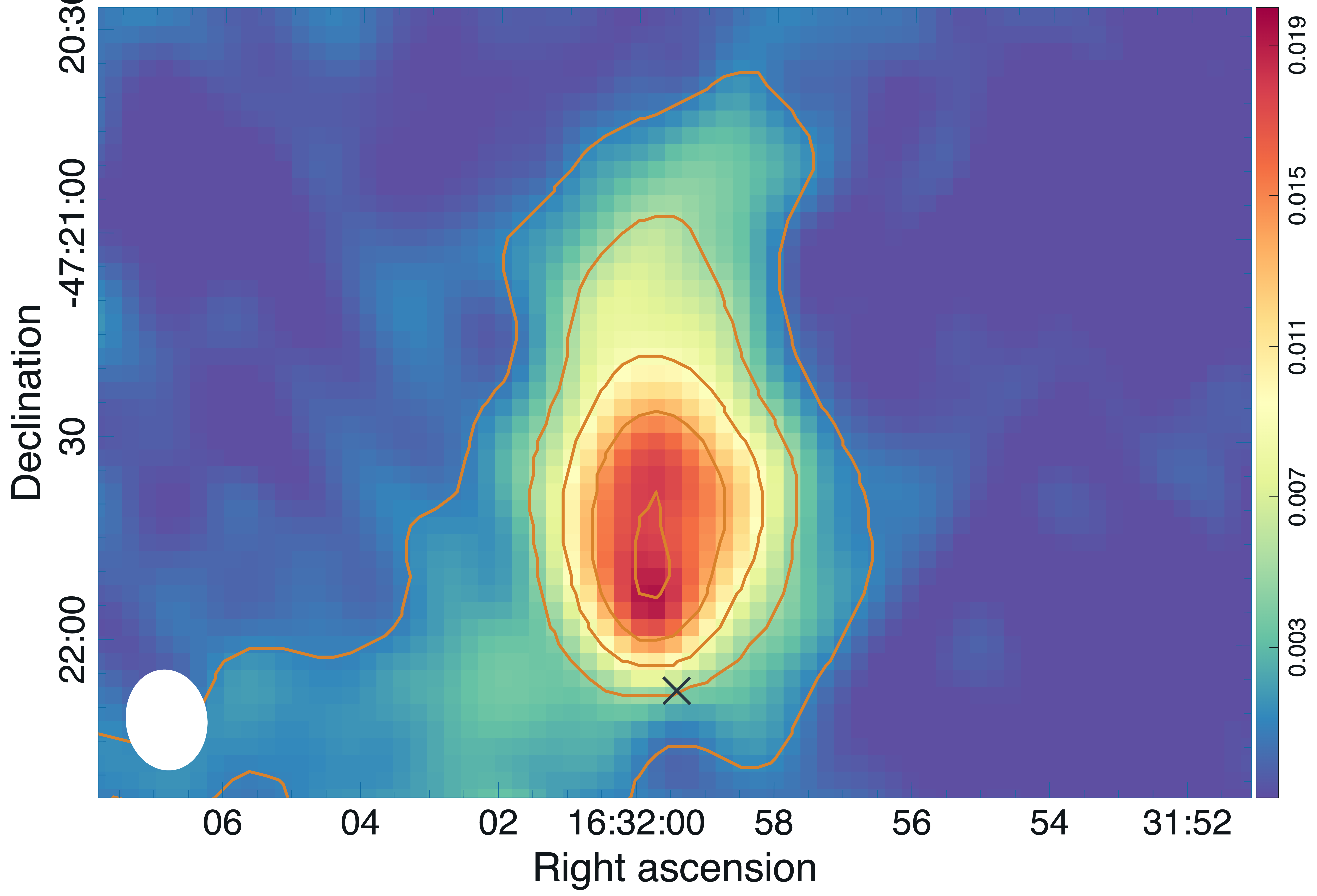}
    \caption{The zoomed-in view of the potential associated PWN enclosed in rectangular region (Fig.~\ref{PWN}). The contours are 1, 5, 9, 13, and 17\,mJy\,beam$^{-1}$. The colour bar is in units of Jy\,beam$^{-1}$, and the beam size is shown in the lower left corner of the image. The black cross shows the position of PSR~J1631--4722.}
    \label{fig8}
    \end{center}
\end{figure}

The current timing of J1631--4722 gives us a pulsar position of RA(J2000)\,= \,16$^{\rm h}$31$^{\rm m}$59.42$^{\rm s}(3)$, Dec(J2000)\,= $-$47$^{\rm o}$22$\arcmin$7.1$\arcsec$(2), which is significantly offset from the extended radio continuum emission (see Fig.~\ref{fig8}). This offset may partly arise from uncertainties in the current timing position caused by pulsar timing noise, and it is likely that the error bars on our current timing position are underestimated. Young pulsars are known to exhibit significant timing noise~\citep[e.g.,][]{hll+05,psj+19}, and our one-year timing baseline is insufficient to resolve the degeneracy between the pulsar's position and timing noise. Based on previous timing studies of young pulsars~\citep[e.g.,][]{psj+19, pjr+20}, a timing baseline of at least five years is expected to allow for a reliable sub-arcsecond measurement of the pulsar's position, providing the precision needed to understand the positional offset between the pulsar and the potential PWN.
%
In the RACS radio continuum image, a cometary tail-like extended emission following the pulsar is observed, which is morphologically a characteristic of PWN. The length of tail-like emission from PWN is $\sim$~73\,arcsec and the structure is pointing away from the geometric centre of SNR~G336.7+0.5 [RA(J2000)\,= \,16$^{\rm h}$32$^{\rm m}$02$^{\rm s}$, Dec(J2000)\,= $-$47$^{\rm o}$19$\arcmin$08$\arcsec$], suggesting that the pulsar is moving away from the supernova explosion site.
%

The distance to SNR\,G336.7+0.5 was reported as 8.7\,kpc \citep{case98}, 9.5\,kpc \citep{stupar07}, and 7.2\,kpc \citep{pavlov13}, all derived using the radio-brightness--to--diameter relation. 
Using the YMW16~\citep{yao17} and NE2001~\citep{ne2001} models for the Galactic distribution of free electrons, we find the distance for a DM of 873\,pc\,cm$^{-3}$ to be 6.8\,kpc and 9.1\,kpc, respectively.
Despite significant uncertainties, the DM-based estimates are consistent with the distances to the SNR\,G336.7+0.5 reported in the literature.
%
If the PWN associated with the pulsar is confirmed, X-ray studies of the PWN could offer additional insights into the distance of this system.
Assuming the true age is similar to the characteristic age of 33\,kyr, we can estimate the transverse velocity of the pulsar using its projected distance from the SNR geometric center. At pulsar distances between 7 and 9\,kpc, the transverse velocity ranges from 175 to 225\,kms$^{-1}$, consistent with previous studies of young pulsar velocity distributions~\citep[e.g.,][]{ydl+21,igo20,hobbs2005}.

PSR J1638--4713 is one of the known handful of highly scattered pulsars~\citep[e.g.,][]{lfd+24,wang23,camilo21}, making it extremely challenging to detect in previous pulsar surveys. This is evident with the fact that the region containing PSR~J1638--4713 underwent pulsar searches as a part of the Parkes Multibeam Pulsar Survey \citep[PMPS;][]{manchester2001} and the High Time Resolution Universe Survey \citep[HTRU;][]{keith10}, both operating at 1.4\,GHz, yet it remained undiscovered.  
%
%
%
Targeted searches at high-frequency offer a promising avenue for discovering more highly scattered pulsars of this type. 
The superior sensitivity and capabilities of the latest generation of interferometers provide all-sky deep radio continuum surveys, such as Evolutionary Map of the Universe with ASKAP \citep[EMU;][]{norris11,norris21}, facilitating the search for diffused and low surface brightness objects in the Galactic plane (e.g., SNRs and PWNe). The radio emission in continuum surveys is time-averaged and independent of pulsed--emission. 
This effectively allows the identification of extreme pulsar candidates (e.g., highly scattered, highly accelerated, and highly intermittent) in the imaging domain, potentially missed in previous pulsar surveys. Subsequently, these candidates can be confirmed through targeted deep observations using wideband systems, such as UWL receiver on Murriyang~\citep{hobs2020}, avoiding the expensive pixel-by-pixel searches at high frequencies.

\section*{Acknowledgements}
We would like to thank the anonymous referee for their feedback that has improved the quality of this work. Murriyang, CSIRO's Parkes radio telescope is part of the Australia Telescope National Facility\footnote{\url{https://www.atnf.csiro.au}} (ATNF) which is funded by the Australian Government for operation as a National Facility managed by the Commonwealth Scientific and Industrial Research Organisation (CSIRO). We acknowledge the Wiradjuri people as the traditional owners of the Observatory site. The ATNF Pulsar Catalogue at \url{https://www.atnf.csiro.au/research/pulsar/psrcat/} was used for this work.
CSIRO’s ASKAP radio telescope is also a part of ATNF. Operation of ASKAP is funded by the Australian Government with support from the National Collaborative Research Infrastructure Strategy (NCRIS). ASKAP uses the resources of the Pawsey Supercomputing Research Centre. Establishment of ASKAP, Inyarrimanha Ilgari Bundara, the CSIRO Murchison Radio-astronomy Observatory and the Pawsey Supercomputing Research Centre are initiatives of the Australian Government, with support from the Government of Western Australia and the Science and Industry Endowment Fund.
The data processing was performed on the OzSTAR\footnote{\url{https://supercomputing.swin.edu.au/ozstar}} national facility at Swinburne University of Technology. The OzSTAR program receives funding in part from the Astronomy National Collaborative Research Infrastructure Strategy (NCRIS) allocation provided by the Australian Government, and from the Victorian Higher Education State Investment Fund (VHESIF) provided by the Victorian Government.

\section*{Data availability}
The observations from Murriyang are publicly
available from \url{https://data.csiro.au/domain/atnf} after an 18-month embargo period. This study made use of archival ASKAP data obtained from the CASDA\footnote{\url{https://data.csiro.au/domain/casdaObservation}}. 


\bibliography{pulsars}
\bsp

\label{lastpage}
\end{document}